\documentclass[10pt,conference,compsoc]{IEEEtran}

\usepackage{cite}
\usepackage{amsmath,amssymb,amsfonts}
\usepackage{algorithmic}
\usepackage{graphicx}
\usepackage{textcomp}
\usepackage{xcolor}
\usepackage{braket}
\usepackage{graphicx}
\usepackage{hyperref}
\usepackage[utf8]{inputenc} %useful to type directly diacritic characters
\usepackage{rotating}
\usepackage{wrapfig}
\usepackage{tabularx}
\usepackage{xspace}
\def\BibTeX{{\rm B\kern-.05em{\sc i\kern-.025em b}\kern-.08em
    T\kern-.1667em\lower.7ex\hbox{E}\kern-.125emX}}

\let\oldcite\cite % Copy \bibitem into \oldbibitem
\renewcommand{\cite}[2][]{\oldcite{#2}}% Redefine \bibitem to only use mandatory arg

\hypersetup{colorlinks=true }

\IEEEoverridecommandlockouts % Don't forget this command!

\newif\ifdraftcolors
\ifdraftcolors

\else

\fi
\begin{document}

\title{Quantum Text Encoding for Classification Tasks \\
\normalsize{Preprint of paper that appeared in International Workshop on Quantum Computing \\ 
IEEE/ACM 7th Symposium on Edge Computing (SEC),
Seattle, WA, December 8, 2022. Pages 355-361. \\
Official version at \url{https://doi.ieeecomputersociety.org/10.1109/SEC54971.2022.00052}
}
}
% \normalsize{
% Status: Abstract accepted for \url{https://iwqc22.github.io/} \\
% Final copy due 2022-08-12 \\
% \bf{Please ask before sharing outside IonQ until this appears on arxiv}}}

\author{\IEEEauthorblockN{Aaranya Alexander}
\IEEEauthorblockA{IonQ, Inc. \\
\texttt{aaranya@ionq.com}}
\and
\IEEEauthorblockN{Dominic Widdows}
\IEEEauthorblockA{IonQ, Inc. \\
\texttt{widdows@ionq.com}}
}

\maketitle

\begin{abstract}
This paper explores text classification on quantum computers. Previous results have achieved
perfect accuracy on an artificial dataset of 100 short sentences, but 
at the unscalable cost of using a qubit for each word. 

This paper demonstrates that an amplitude encoded feature map combined with a quantum support vector 
machine can achieve 62\% average accuracy predicting sentiment using a dataset of 50 actual movie reviews.
This is still small, but considerably larger than previously-reported results in quantum NLP.
\end{abstract}


%%%%
\section{Introduction, Motivation, and Outline}

Quantum natural language processing (QNLP) is a new field, which in 2022 is theoretically advanced and nascent
in quantum implementation.
Quantum mathematical models have been used since the early 2000's in language-related fields such as information retrieval 
\cite{rijsbergen2004geometry,widdows2004geometry} and cognitive science \cite{aerts1993quantum}, and by 2010
deliberately quantum-theoretical approaches to combining logical and distributional semantics led to the development
of the Distributed Compositional Categorical model of \cite{coecke2010distributional} (subsequently shortened to DisCoCat).
This and related techniques including tensor networks, Frobenius algebras, and density matrices have been used
to design systems that have demonstrated quantitative successes on various language tasks 
(for surveys see \cite{widdows2021quantum,guarasci2022qnlp}). As a source of scientific models and theories that
have successfully been implemented, QNLP has become quite advanced.

By contrast, only in the past two or three years has it been possible to run QNLP programs on actual quantum computers.
In 2020, experiments were run using 6-qubits \cite{lorenz2021qnlp}, demonstrating successful compilation of
short sentences into quantum circuits to give 83.3\% classification accuracy. Since then, less sophisticated but more
accurate classification results have been obtained at the cost of using more qubits (8 or 11) \cite{widdows2022near},
and this work has also demonstrated small examples of circuits used as part of natural language generation and disambiguation.
So QNLP is nascent, in that the programs implemented on quantum computers are very small and young.

Similar situations are typical today in quantum information processing: the field has produced striking theoretical 
results since the 1980s and 1990s, and quantum computers that can implement these ideas are only just becoming available.
The pace of development is fast --- for example, at IonQ alone, systems have progressed from the 11 qubit machine evaluated
by \cite{wright2019benchmarking} to regularly running jobs with 20+ qubits \cite{ionq2022AQ}. The key scaling property of quantum memory
is that the number of variables doubles with each additional qubit: so in theory, 10 qubits corresponds to a kilobyte, 20 to
a megabyte, and 30 to a gigabyte, and it is easy to see that with 50 to 100 addressable qubits, the gap in scale 
between quantum and classical NLP could be closed.

However, adapting even relatively simple NLP models to use these resources requires work.
 One strategy would be to wait for theoretical systems like the `bucket brigade'
protocol of \cite{arunachalam2015robustness} to be fully available, at which point implementing memory-intensive processes
should be much simpler --- but many opportunities may be forgone in the meantime, including the opportunity
to influence the design of memory access, and to deliver useful intermediate-scale systems. 
An alternative more hands-on approach is to try and get the best results we can with current systems, to 
use this process to inform the design of useful intermediate-scale systems as soon as possible, and to cooperate
directly with hardware engineering to enhance both machines and applications together.  
So far this approach has demonstrated that accurate
classification can be performed using quantum hardware on a very small dataset, but with the memory requirement
of at least one qubit for each salient word \cite{widdows2022near}. 
This method would not scale to a significant vocabulary 
without thousands of qubits --- but if a more space-efficient method can be used to store and combine word-topic weights,
the memory requirement could be much smaller. 

This motivates the search for more space-efficient text encoding techniques to use in QNLP tasks, which is the topic
of this paper. 

This paper outlines the results of the preliminary study into space-efficient quantum encodings for binary text classification.
The new research focuses on the implementation of encodings to enhance the Quantum Support Vector Machine (QSVM) classification method.
First, the performance of different quantum classifiers are revisited, and the theory behind the quantum enhancement to the QSVM is 
explored. Further sections compare results across different encodings of text data in both quantum and classical methods. 

%Finally, 
%groundwork conclusions are drawn concerning quantum classification specifically for text data, and its potential applications for 
%medium-term NLP goals. \par

% This paper reports on early efforts and results in such an endeavor. 
% The goal is to investigate 
% methods to encode text more efficiently, to be able to scale up to larger datasets while preserving 
% the quality of results already demonstrated on smaller and more artificial challenges.
% The paper is organized as follows. Section \ref{sec:qml_background} outlines classification
% approaches in the context of quantum machine learning. Section \ref{sec:quantum_classification} reviews
% how these have been applied to text classification so far. 

%%%%%%
\section{Background: Quantum Machine Learning and Classification}
\label{sec:qml_background}

\subsection{Supervised and Quantum Machine Learning}

The experiments in this paper follow the pattern of many machine-learning approaches to classification tasks.
The task is to determine whether a piece of text is about a particular topic (e.g. {\it food} vs. {\it computing}),
or demonstrates a particular sentiment (e.g. {\it good} vs. {\it bad}). 
The system is given training examples where the correct
label is provided, and then evaluated by seeing how often it assigns the correct label to test data not used in training.
The use of annotated training and test data makes this a supervised learning approach \cite[Ch 1]{geron2019hands}.

Quantum machine learning has risen in potential in recent years due to development of quantum algorithms 
with space-efficiency and speedup compared to their classical counterparts. 
As described in \cite{schuld2021machine}, `quantum machine learning' could refer to the use of quantum models 
on classical hardware with classical data, and this category covers most of the successes of QNLP to date. The 
new area for quantum computing is the opportunity to apply quantum processing to classical data, and the experiments in this
paper fall into this category.

% A subsection of machine learning concerns natural language processing (NLP), a core element to creating fully intelligent systems. 
% Specifically, quantum natural language processing (QNLP) has gained traction by exploiting the similarities between grammar and language composition with quantum data structures and algorithms \cite{lorenz2021qnlp}. The promise of QNLP and its most recent advancements are introduced in \cite{widdows2022near}, focusing on primary tasks of topic classification and bigram language modeling. \par

% The complexity of language alone means that QNLP systems should extract meanings of text depending on definitions of individual words, contextual indicators from surrounding words, and the arrangement of such words in sentences\cite{widdows2022near}. Nuances and interconnectedness of language began a push towards using machine learning concepts like neural networks, vectors and linear algebra to represent text features. The largest factors limiting classical natural language processing are the ever-growing datasets and subsequent difficulty in pattern fitting, data storage and processing time \cite{guarasci2022qnlp}. Quantum systems have become attractive due to the possibility of performing accurate data manipulations in conjunction with space efficiency of the encoded quantum state. \par

% Text classification tasks performed fully with quantum concepts and hardware remain limited in practice \cite{guarasci2022qnlp}\cite{widdows2022near}. 

Quantum algorithms make necessary trade-offs between space-efficiency, accuracy, and code complexity. % \cite[Ch 3, 4]{schuld2021machine}.
For example, encoding data densely in a quantum state may bring difficulty in extracting information without disturbing the system, and often requires complicated circuits to perform data manipulation. However, space-efficient quantum algorithms in the realms of image classification and statistical datasets have shown to be promising in limiting extreme trade-offs between efficiency, accuracy and space \cite{li2021besiii}\cite{shan2022cancer}. 
This also emphasizes the need for investigation of such methods in text datasets.

%%%%
\subsection{Featurizers and Classifiers}

Many machine learning systems for classification today follow the pattern of featurizing then classifying.
Various featurizers and classifiers are used in this paper, and describing them in the way can help to understand
system architectures more easily.

\subsection{Featurizers}

A featurizer (sometimes called an encoder) takes an input such as an image or text file and maps it to a list 
of weights of salient features, which can be seen as a vector. 
These could be explicit features, such as the proportions of red, green, and blue
in an area of an image, or implicit learned features, such as coordinates produced by a principal component analysis.
There are many ways to map text datasets to feature vectors, ranging from counting the number of times each word occurs
in each document (used since at least the 1960's), to using a neural network trained to map text to vectors (with a 
goal of predicting as many missing words as possible). The family of neural network methods has become large and 
includes word-based models such as Word2Vec \cite{mikolov2013efficient} and contextual methods such as BERT \cite{devlin2018bert}
that featurize and combine fragments of words. Such text-to-vector featurizers are often referred to as {\it encoders},
and their output vectors are often called {\it embeddings}.

\subsection{Classifiers}

A classifier takes an input and maps it to a class label (or several). That is a very general description, and could
be implemented in many ways. The benefit of the featurizer / classifier patter is that the input to a classifier
is typically reduced to vectors: in terms of types and interfaces, instead of mapping {\it anything} to a class label,
a classifier can be implemented that maps a vector of floating point values to a class label.

%%%
\subsection{Datasets Used in this Paper}

\subsubsection*{Lambeq Dataset}

The first dataset used in this work is the 70 training and 30 development and test sentences used 
for topic classification experiments by \cite{lorenz2021qnlp},
and subsequently released as an open source package called {\it Lambeq} \cite{kartsaklis2021lambeq}. 
The sentences are artificially generated to use a small fixed vocabulary, to follow predictable syntactic patterns,
and each comes with a binary topic label, `0' indicating {\it computing} and `1' indicating {\it food}, as in these examples:

\medskip
\begin{quote}
\small
\texttt{1  man prepares meal .}

\texttt{0  skillful woman debugs program .}
\end{quote}
\medskip

\noindent
The vocabulary used is too small to be representative of natural language, but 
this is still a useful and appreciated contribution to the QNLP
community, because it enables use to quantitatively evaluate results on a dataset that is small enough 
to be loaded on today's quantum hardware.

\subsubsection*{IMDB Dataset}
The second, more complex set is taken from 50,000 archived IMDB movie reviews to be classified as either positive or 
negative reviews \cite{maas2011imdb}. The average number of words in a review is between 228 and 229.
Sentiment classification for this dataset is more difficult, because it is real user-generated text
with varying degrees of good and bad, and many different aspects of a movie, some of which might
be described positively and others negatively. 
Nonetheless, the reviews are published in positive and negative
directories, and thus each review comes with a binary sentiment label.

\subsection{Text Preprocessing}

The datasets used are all English language and were preprocessed using simple methods.
The Lambeq dataset can be perfectly tokenized just by splitting on whitespace characters, while
the IMDB dataset has more of the vagaries of normal natural language. All of the experiments
in this paper that used the IMDB dataset used a version of Word2Vec \cite{mikolov2013efficient}
to encode words as vectors, which also provides basic elements of tokenizing and normalizing English, 
in this case including splitting on whitespace, removing punctuation, though not automatically lowercasing.
 
%%%
\section{Bag of Words Approach}

\begin{figure}[ht]
    \centering
    \includegraphics[width=0.47\textwidth]{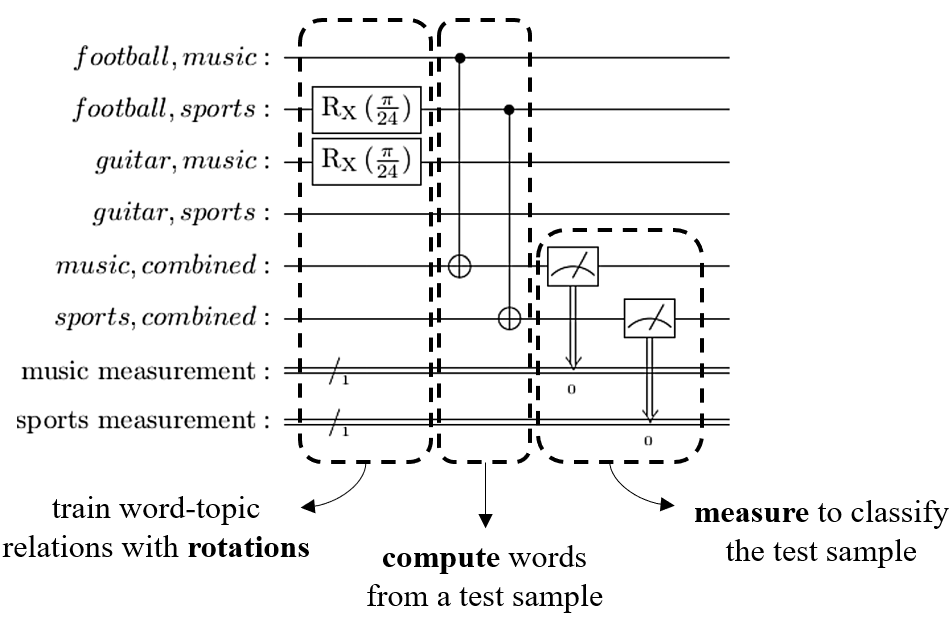}
    \caption{QBoW trained on two words, football and guitar, into topics, sports and music. Classification of a sample containing the word football \cite{widdows2022near}.}
    \label{fig:qbow}
\end{figure}

One of the simplest families of classification methods are Bag of Words (BoW) approaches. In such methods, the features
for each word are just added together --- other conditional dependencies between features (such as those arising from
word-order) are ignored.
For a set of training documents with a known classification, the BoW classifier keeps a score per encountered word, where the score is 
proportional to the frequency of a word in each topic. These scores can be stored classically in a word-topic matrix. For new training 
documents, the classifier sums the scores of the words in the training set, and the topic with the highest score is deemed the class of 
the document. In the naive quantum implementation (Figure \ref{fig:qbow}), the relationship between a word and a topic is encoded with 
single qubit rotations in the training phase. Then, a common phase adding circuit is performed on “combined” topic qubits to sum the 
scores of words, and measured to classify the sample. \par

This can be seen as a featurizer / classifier pattern, with two very simple parts. The featurizer uses classical memory
to map a text to vector for each topic, by mapping each (word, topic) pair to a particular qubit. The classifier adds the coordinates for each qubit belonging to the corresponding topic into a single combined value for that topic.

This quantum BoW circuit achieved 100\% accuracy for the Lambeq dataset, the first such result reported on a quantum computer. 
However, it is extremely unscalable as it requires a qubit per (word, topic) combination, and additional qubits to hold each overall topic score.
% Dom: I removed this because it gets more complicated - we could use different rotation angles for different words, but how would we store / record the difference, or we could use a kind of clock protocol with several dials with different periodicity - again, possible but complicated. 
%Encoding with single qubit rotations is impossible for large datasets due to the periodicity of the rotation angle, and the simple adder circuit also fails with increased vocabulary. 
It follows that a goal in overcoming the bottleneck of quantum scalability is to improve the qubit encoding of the words in addition to the means of classification.

%%%
\section{Support Vector Machines}

\begin{figure}[ht]
    \centering
    \includegraphics[width=0.47\textwidth]{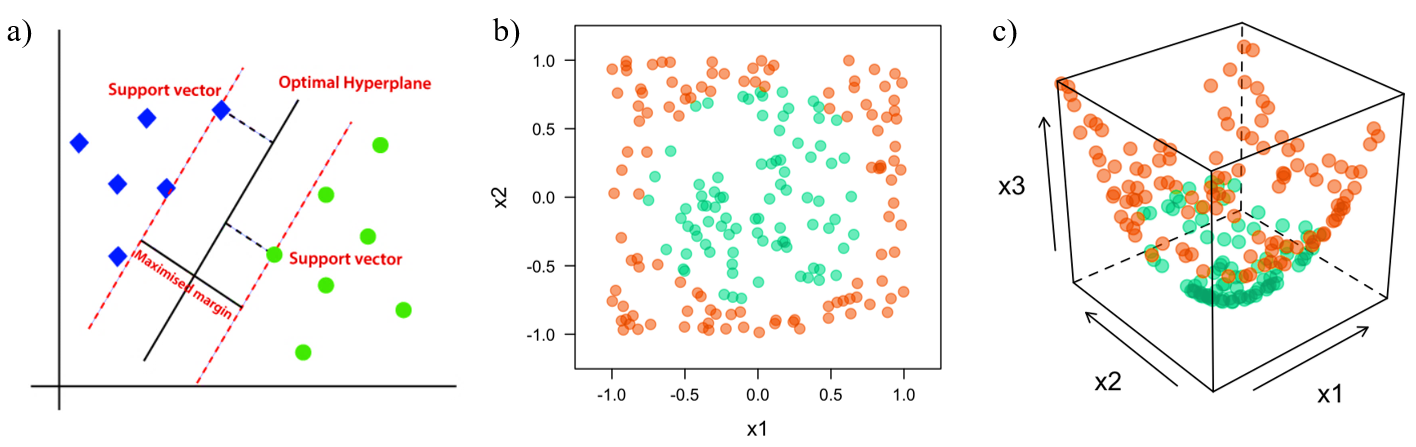}
    \caption{a) Well-separated, linear feature space \cite{salunkhe2021svm} b) Low margin, non-linear boundary \cite{boehmke2019HandsOnML} c) Multi-dimensional boundary after feature space enlargement of b) \cite{boehmke2019HandsOnML}}
    \label{fig:fspace}
\end{figure}

Support Vector Machines (SVMs) classify arbitrary vectors by mapping training vectors to higher-dimensional spaces, 
and determining a class boundary which is used to classify new test vectors \cite{boehmke2019HandsOnML}. 
Figure \ref{fig:fspace} demonstrates three important cases of vector datasets that can be helped by the SVM. In a well-separated, linear, 
binary problem, classical fitting methods can confidently find an optimal hyperplane between the classes. However, as complexity of the 
data increases, exact nonlinear boundaries with large margins are difficult to achieve. Thus, the SVM first 
maps the data onto an enlarged feature space with a feature map, so the classes may be more easily distinguished. After application of 
the feature map, a kernel matrix is generated to store the relationship of support vectors with each other before being passed into the 
SVM for fitting. \par

Quantum SVM classifiers have been proposed to harness higher-dimensional Hilbert spaces as feature spaces, and use known quantum computation methods to calculate the kernel\cite{havlicek2019supervise}. 
The division of labor between classical and quantum processes is described in detail by \cite[Ch 6]{schuld2021machine}.
The key insight is based on recognizing a similar structure between kernel methods and quantum 
processes. Kernel method bring a key optimization to the process by computing a predicted similarity
between two high dimensional vectors by applying an appropriate kernel similarity function to lower-dimensional 
counterparts, thus avoiding the need to calculated the mapping into the higher dimensional space explicitly.
Quantum circuits work well for exploring this higher-dimensional space, finding the pairwise similarities
between different training instances that are then used as entries in the kernel matrix, which can then be used
to train an SVM classifier using entirely classical computing.
As such, the QSVM describes a hybrid quantum-classical method, where classical vectors are mapped to quantum states with a quantum feature map. 

The quantum kernel can be efficiently calculated from the encoded quantum vectors, and then passed into the classical SVM. 
% For two vectors $\vec{x}$ and $\vec{z}$ and a quantum feature map $U_{\phi}$, the kernel element 
% $\kappa(\vec{z}, \vec{x})$ in Equation (1) is estimated by the circuit in Figure \ref{fig:kernelcirc} \cite{havlicek2019supervise}. 
The feature map is applied to the $\ket{0}^n$ state as a function of $\vec{x}$, 
and then its conjugate is applied as a function of 
$\vec{y}$. 
The kernel value is estimated by executing this circuit over a number of shots, R. The fraction of occurrences where the 
`0' string is measured corresponds to the estimated kernel value. \par

\begin{equation}
    \kappa(\vec{y}, \vec{x}) = |\bra{0^{n}}U_{\phi(\vec{y})}^{\dag}U_{\phi(\vec{x})}\ket{0^{n}}|^2 = |\braket{\vec{\phi({y})}|\vec{\phi({x})}}|^2
\end{equation}

\begin{figure}[t]
    \centering
    \includegraphics[width=\linewidth]{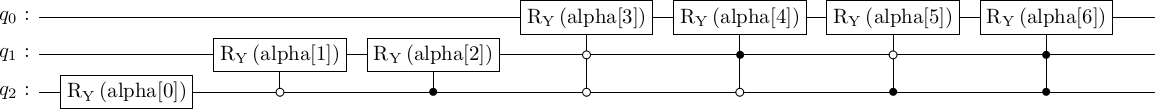}
    \caption{Amplitude encoding circuit with 3 qubits}
    \label{fig:amplcirc}
\end{figure}

\begin{figure}[t]
    \centering
    \includegraphics[width=0.8\linewidth]{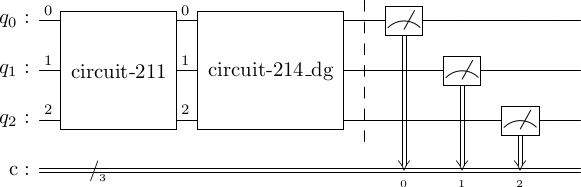}
    \caption{Kernel value estimation circuit with 3 qubits}
    \label{fig:kernelcirc}
\end{figure}

For example, a 7-dimensional amplitude feature map can be encoded with the 3-qubit circuit of
Figure \ref{fig:amplcirc}. Such an operator for $\vec{x}$ is then concatenated with a corresponding 
inverse operator for $\vec{y}$, and the estimated kernel value is given by the proportion of $\ket{000}$
outcomes measured. (These circuits were constructed and run using the Qiskit package \cite{qiskit2021textbook}.)

%The QSVM appears to be a promising approach for text classification, where it would be ideal to access as many feature dimensions as possible to capture details in language features. 
The QSVM has been successfully implemented for classification of non-text datasets, 
with comparable accuracy to its fully classical analog \cite{shan2022cancer}\cite{li2021besiii}.
The main limitation of the QSVM concerns the design of the quantum feature map. For an n-qubit feature map, the accessible enlarged feature space is of size $2^n$, which is exponentially greater than what is possible with n-bits of classical space\cite{havlicek2019supervise}. This quantum map must also manipulate the feature vectors so that the kernel matrix elements represent a classifiable relationship between vectors, and have reasonable complexity to be executed over many shots. \par

\subsection{Simple QSVM for Text Classification}

% In order to use the vector classification techniques explored further in this section, the text samples must have efficient numerical representations that can be encoded into a quantum state. The Word2Vec method developed by \cite{mikolov2013w2v} does so efficiently via neural network; words with contextual or logical similarity are reflected by their corresponding Word2Vector’s cosine similarity. The package constructs the float vectors to a certain length known as the embedding dimension, where a larger embedding dimension allows for increased patterns and nuances to be represented between words \cite{mikolov2013w2v}. The text datasets researched were first converted to Word2Vec representations then used as inputs to the QSVM feature maps.

In an initial experiment using the Lambeq dataset, a QSVM classifier was coupled with the use of Word2Vec embeddings as features (described above).
Sentence vectors are generated for each sentence by computing a Word2Vec embedding for each word, then averaging all the words in a sentence. This part is all done classically. These feature vectors were then passed to the QSVM for training, and used by the QSVM
for classifying new sentences. Using 8 dimensions and one qubit for each dimension, an accuracy of 90\% was achieved 
\cite{widdows2022near}. The quantum memory requirement is thus one qubit per dimension rather than one qubit per word. 
As a general rule, embedding dimensions tend to be a few hundred, ranging from default values of 300 for Word2Vec and 768 for BERT.

% and representative vectors for each review are computed by 
% tokenizing the review text, then averaging the Word2Vec embeddings of all the words in a review ($\sim$100-200 words).
% References to the IMDB dataset concern a random selection of movie reviews for train and test data,
% and reference to the Movies dataset regards a contrived 
% dataset. The Movies dataset takes reviews from the minimum number of movies to meet the desired test and train set size, in an attempt to 
% reduce niche vocabulary and improve the quality of the training.

\section{Text Encoding in Feature Maps} 

\subsection{One-Hot Encoding Feature Map}

The Pauli Gate feature maps are the most widely used in QSVM experiments for their low depth and adequate complexity. The second-order Pauli-Z evolution circuit (ZZ Feature Map) performs a non-linear mapping from $n$ features to $n$ qubits, analogous to a “one-hot” 
encoding \cite{orazi2022qml}. In this case, it uses minimum space-efficiency achievable for a quantum feature map ($n$-to-$n$ mapping), 
but obtains a quantum advantage due to the inability to simulate the ZZ feature map classically at larger scales.

\begin{figure}[ht]
    \centering
    \includegraphics[width=\linewidth]{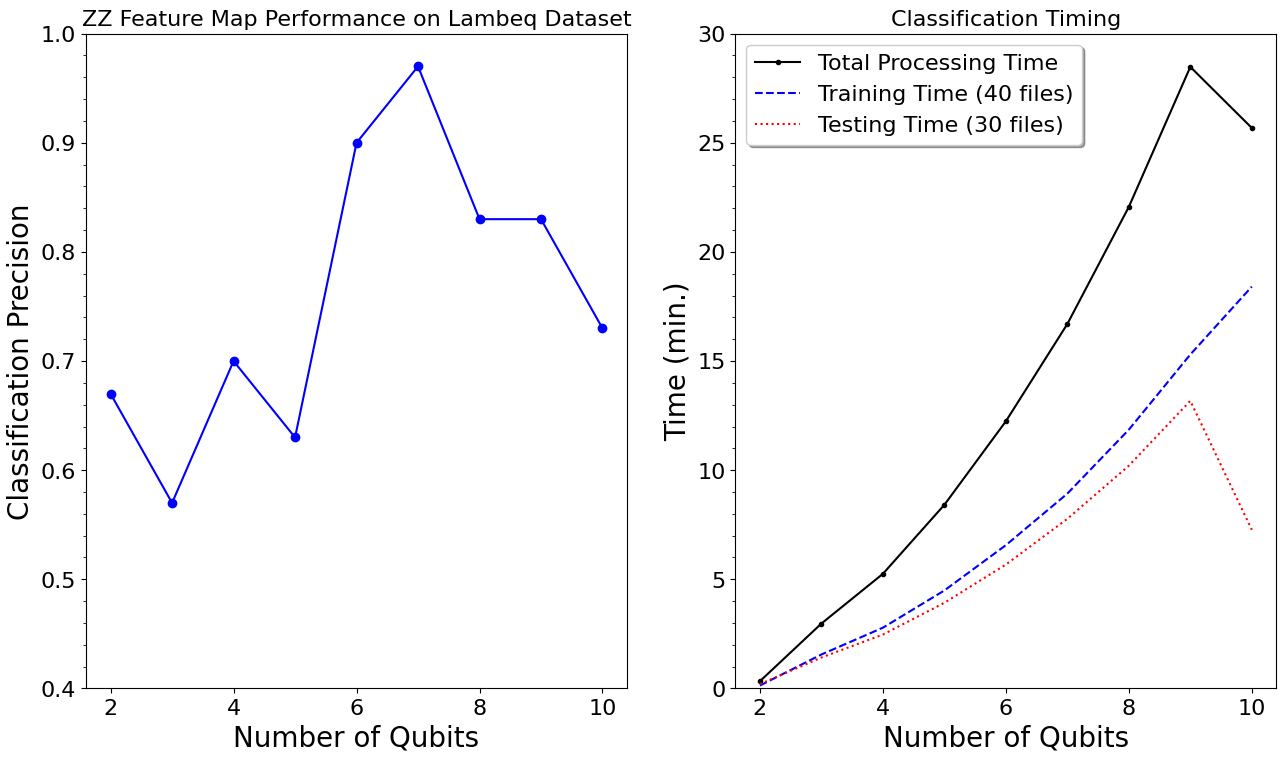}
    \caption{Classification Results on Lambeq Dataset Using One-Hot Encoding Feature Map}
    \label{fig:lambeqZZ}
\end{figure}

An initial smaller-scale experiment was carried out using the IonQ simulator.
For the Lambeq dataset, the classification accuracy peaked at 97\% for 7-dimensional feature embeddings (Figure \ref{fig:lambeqZZ}). 
The exponential increase in processing time with the number of qubits is infeasible for realistic NLP data; seven qubits for each inner product calculation scales poorly given the simplicity and low size of the Lambeq dataset compared to real text datasets. \par

%%%
\subsection{Amplitude Encoded Feature Map}
Quantum computing is appealing for several applications due to the potential dense encoding schemes for data into a qubit. Recent work in theoretical formalisms for dense encodings and computations are abundant, ranging from amplitude encodings to simultaneous amplitude and phase combined states. 
Most encoding schemes have few low-depth state preparation methods, although it is a primary focus of current research. Moreover, the ability to extract or manipulate encoded features to perform a text classification task becomes more complicated the denser the encoding. \par

To investigate the performance of dense encodings in QNLP classification, this work takes a baseline approach with the simplest amplitude encoding scheme for a binary classification task (Equation \ref{eqn:2}). \par 

\begin{equation}
    \label{eqn:2}
    \ket{word} = p_{1}\ket{class_1} + p_{2}\ket{class_2}
\end{equation}

Building off of the one-hot encoding approach, it is possible to densely encode the feature elements to represent an n-dimensional Word2Vec vector in \(\log_{2}(n)\) qubits. Using the feature vector encoding as a feature map focuses specifically on developing an idea of text classification performance with a full dense encoding implementation. \par

To map the Word2Vec vectors to the quantum state in Equation \ref{eqn:4}, the divide and conquer state preparation method developed by \cite{araujo2021divide} was implemented. This amplitude encoding feature map consists of a series of controlled-Y rotations applied on the 
\(\ket{0^{\log_{2}(n)}}\) state. Execution of this feature map for kernel estimation will be comparable to a classical linear kernel that takes the dot product of the each pair of vectors. 
It is emphasized that results from this map may not provide a quantum advantage, but acts as a means to draw conclusions on classification of densely encoded text data. \par

\begin{equation}
    U_{\phi(\vec{x})} = CR_{y}(f(x_8))...CR_{y}(f(x_1))
\end{equation}
\begin{equation}
    \label{eqn:4}
    U_{\phi(\vec{x})}\ket{000} = x_1\ket{000} + x_2\ket{001} ... + x_8\ket{111}
\end{equation}

\section{Classification Results}

\begin{figure}[ht]
    \centering
    \includegraphics[width=0.47\textwidth]{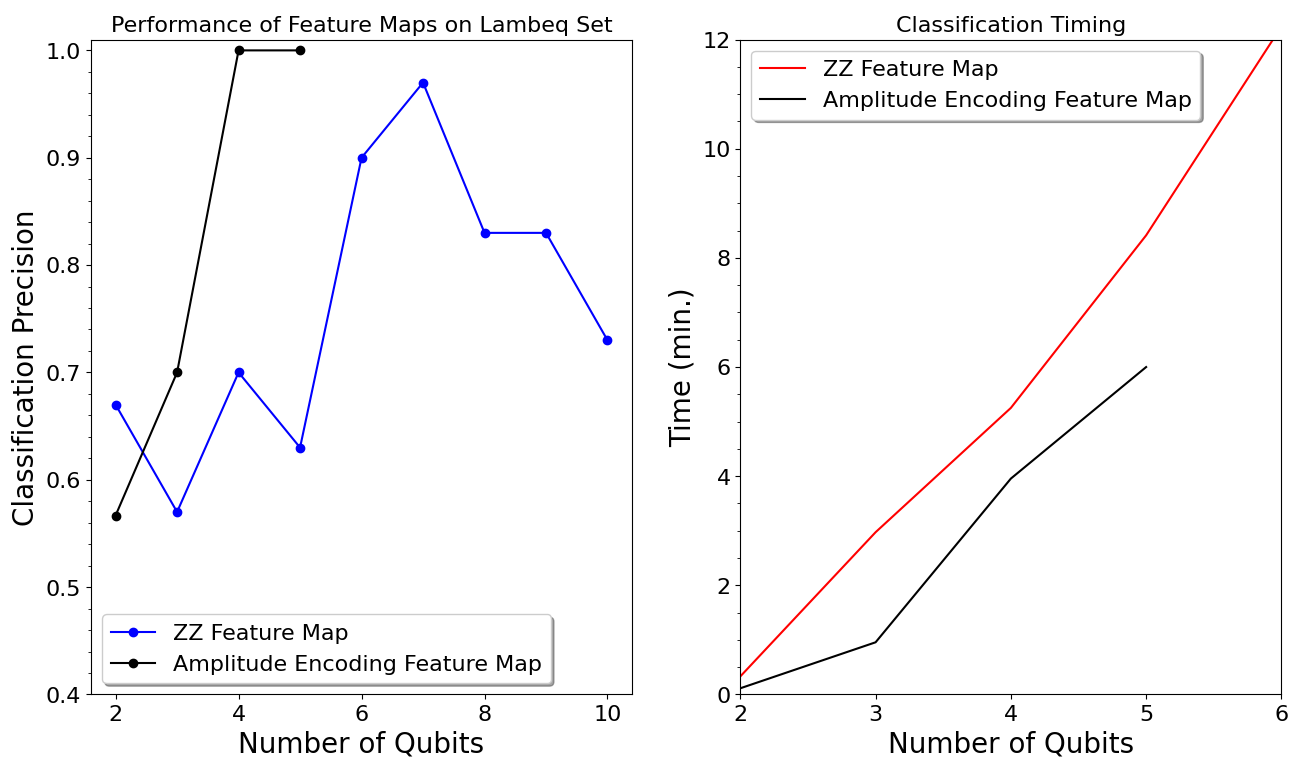}
    \caption{Comparison of Feature Maps on Lambeq Set}
    \label{fig:lambeqamp}
\end{figure}

The results described here were obtained using the IonQ simulator.
The results of the encoded feature map in the QSVM are outlined in Figure \ref{fig:lambeqamp}, for the Lambeq dataset. The amplitude encoding not only exceeds the classification accuracy of the ZZ feature map, but it also reached 100\% classification in just four qubits and lower processing time. \par

\begin{figure}[ht]
    \centering
    \includegraphics[width=0.47\textwidth]{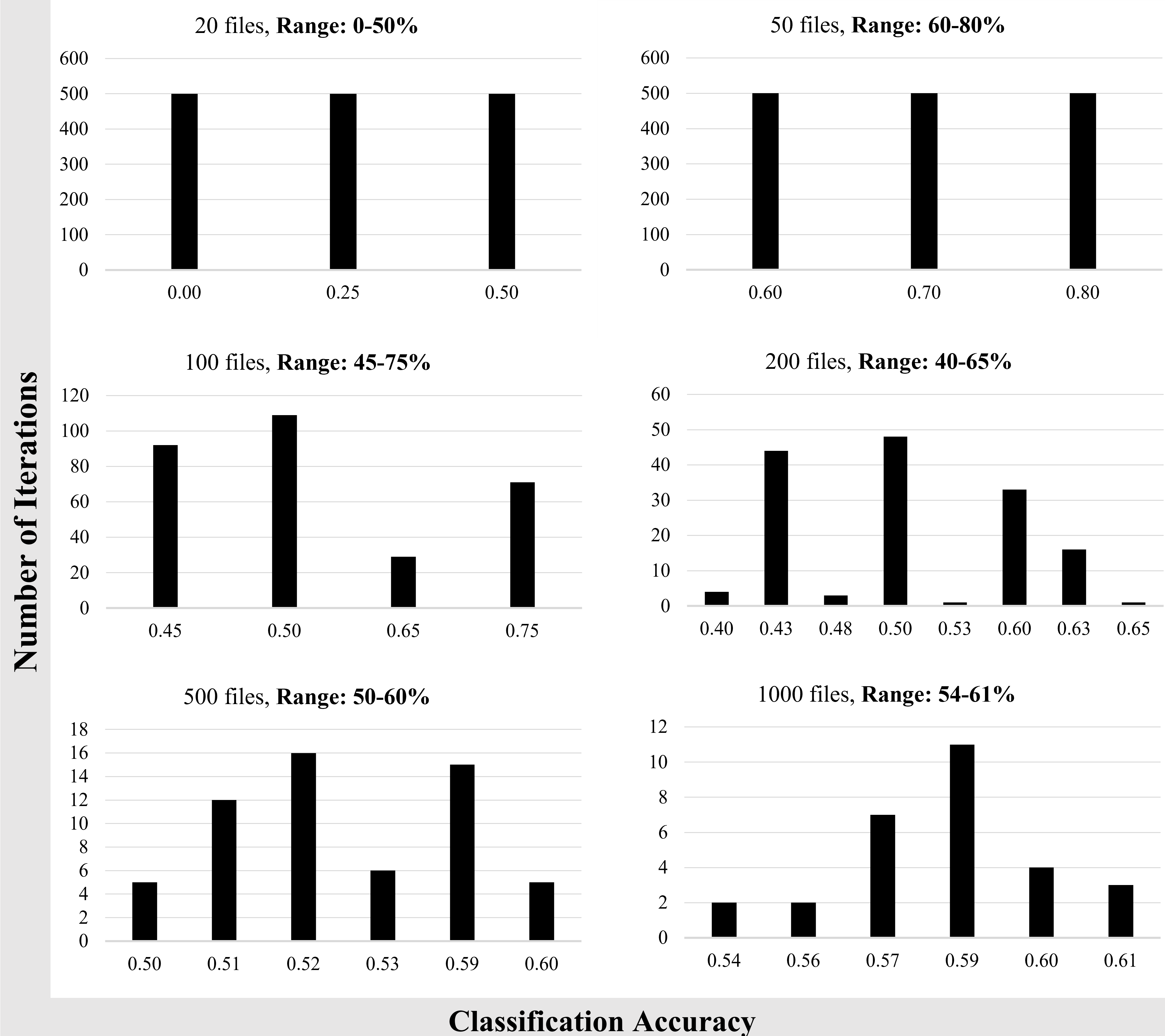}
    \caption{CSVM Performance Distributions using IMDB Dataset}
    \label{fig:classdist}
\end{figure}

Performing further testing with the IMDB dataset, the classification accuracy varied considerably with each iteration. There remained cases with both the ZZ feature map and the amplitude encoded map where classification did not improve at all with increased features, where some sets peaked at 75\% accuracy for three or four qubits. To corroborate the poor performance, analysis of the IMDB random dataset with the classical SVM (CSVM) and Bag of Words was performed with over several iterations and varied file size. Both failed to achieve greater than 80\% accuracy and overall achieved under 60\% accuracy for datasets with greater than 100 files (Figure \ref{fig:classdist}).\par

% and reference to the Movies dataset regards a contrived 
% dataset. The Movies dataset takes reviews from the minimum number of movies to meet the desired test and train set size, in an attempt to 
% reduce niche vocabulary and improve the quality of the training.

\begin{figure}[ht]
    \centering
    \includegraphics[width=0.47\textwidth]{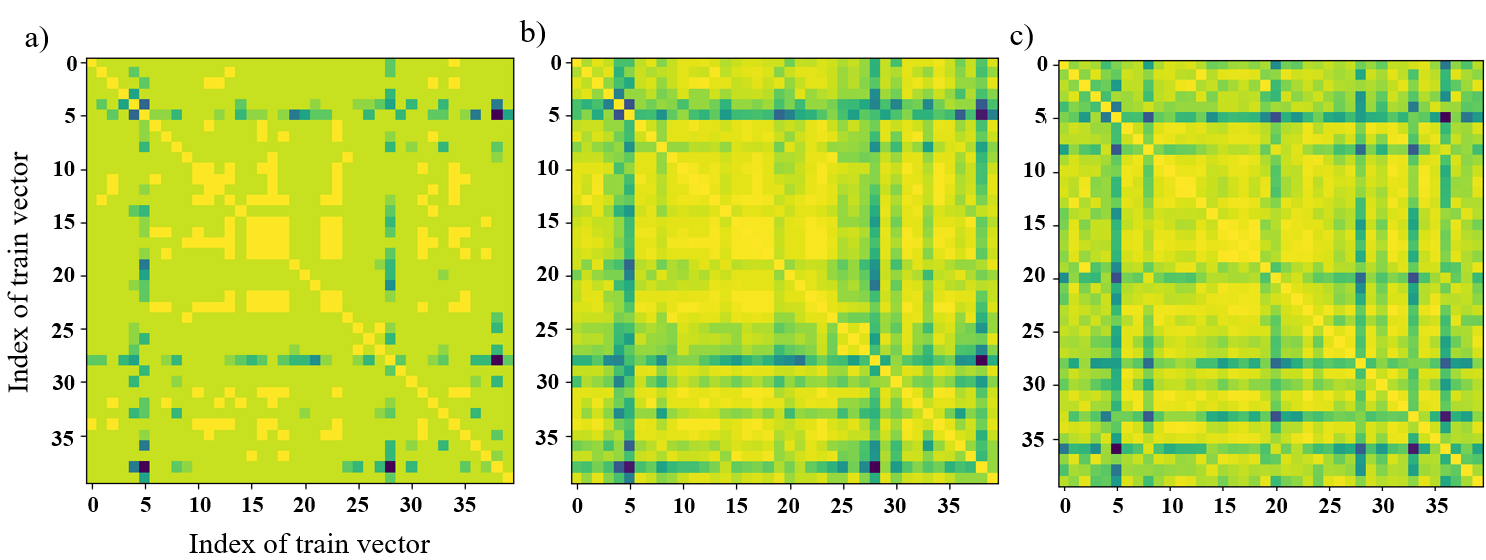}
    \caption{\centering Comparison of a) 1000 shot kernel, b) 10 000 shot kernel with c) true kernel.}
    \label{fig:kernel}
\end{figure}

To model a more targeted scenario, smaller Movies collections were made by 
taking reviews from the minimum number of movies to meet the desired test and train set size, in an attempt to 
reduce niche vocabulary and improve the quality of the training.
The Movies dataset filtered for smaller training and testing vocabulary achieved more consistent trends with different sizes of file sets. 
Improvements in classification accuracy merited increase in kernel estimation shots; the effect of the different shot numbers on the 
kernel accuracy is shown in Figure \ref{fig:kernel}. \par

For 50 files, 10,000 shots were sufficient to achieve decent results, where the systems were trained using 40 files, and tested with 10 files. Average performances for the CSVM and two QSVM maps are outlined in Table \ref{tab:1} for 20 different 50-file samples. The Amplitude Encoded QSVM correctly classified up to 8, 9 and 10 test files on select samples using three, four and five qubits respectively. This is a leading result for classification of real case text samples, of such a feature space, and of this 
complexity. Over all of the samples, a high classification score (70\% or higher) was attained 26\% of the time for both QSVM maps, and 36\% of the time for the CSVM. Moreover, the Amplitude Encoded feature map outperformed the alternatives 80\% of the time when it attained a high classification score. This indicates that the densely encoded feature map is often the best choice for the select samples it performs well on.\par

\begin{table}[ht!] 
    \centering
    \caption{Classification Results for 50 files of the Movies set (10 000 Shots, averaged over 20 samples)}
    \label{tab:1}
    \begin{tabularx}{0.47\textwidth} { 
  | >{\centering\arraybackslash}X 
  | >{\centering\arraybackslash}X 
  || >{\centering\arraybackslash}X
  | >{\centering\arraybackslash}X | }
        \hline
        Word2Vec Embedding Dim. ($n$) & CSVM & ZZ using $n$ qubits & Amp. Enc. using $n$ qubits   \\ [0.5ex] 
        \hline\hline
        2   & 0.579$\pm$0.124           & 0.526$\pm$0.156           & 0.563$\pm$0.122  \\
        3   & \textbf{0.615$\pm$0.160}  & 0.563$\pm$0.169           & 0.537$\pm$0.153             \\
        4   & 0.568$\pm$0.184           & \textbf{0.568$\pm$0.152}  & \textbf{0.621$\pm$0.132}    \\
        5   & 0.611$\pm$0.180           & 0.574$\pm$0.200           & 0.579$\pm$0.164             \\ [1ex] 
    \hline
    \end{tabularx}
\end{table}

\section{Discussion}
 
The results on the Lambeq set are promising for the functionality of incorporating denser encodings into QSVM feature maps. 
There are several key factors behind the overall poor accuracy on the IMDB reviews, first and foremost being that the reviews were real 
text samples incorporating sentiment and colloquial language structure. Conversely, the Lambeq dataset uses controlled skeleton sentences 
with basic grammar, and 200x lower word to sentence ratio than the reviews datasets. Other variational-based quantum methods, like a  
quantum self-attention neural network from \cite{li2022qnn} report similar trends with sentiment classification, and word to sample 
ratios. Such a neural network achieves up to 85\% classification on a separate IMDB sentiment analysis set with no more than 12 words per 
sample, and 100\% reached only on synthetic datasets. \par

Further, QSVM kernel estimation changes the results with several iterations. This is an expected result due to the probabilistic nature of performing computations and measurements with quantum circuits. It introduces more questions with regard to the feasibility of using delicately encoded quantum states, and the necessary shots needed to achieve accurate kernel estimates. The shot increase for the Movies dataset with only 50 files achieves a decent result and processing time, but this cost may become more apparent as quantum text datasets become larger. We can compare the work of \cite{maheshwari2022vqc} that investigated using amplitude encodings in a variational quantum circuit for binary classification. 
Accuracies of 75\% and 67\% were reported for datasets on diabetes patient specifications and sonar signal data, respectively. Such results were achievable with five iterations of the feature map and 100 epochs, and still under-performed compared to the traditional variational quantum classifier. This suggests that competitive results coming from densely encoded data suffers still from increased complexity and reduced accuracy across many datasets.
\par

The non-linear trends in embedding size to classification accuracy highlights the potential unpredictability of classifying natural language datasets versus artificial or non-text datasets. 
%There is a possibility that an optimum number of features in the Word2Vectors exists, that when mapped to the quantum space, generates better margins for a classification boundary. The Word2Vec package does not incorporate known classes of the samples into the embeddings, and in some cases the features it chooses to distinguish samples with may not translate to a good classification boundary. 
Optimum performance may be unique to specific text samples and datasets, as well as the type of task.

QNLP research is at the beginning of even generating these questions, and the varied results from 
each refined dataset is encouraging to develop a systematic approach to improving quantum classification further. \par

%%%%%%%%
\section{Conclusions and Further Work}

The intricacy of completing quantum text classification tasks stems from the nuances of natural language,
and the complicated challenge of encoding this in quantum memory.
The results demonstrated in this paper are a preliminary attempt to investigate the functionality of denser encodings 
in QSVMs, and the connections 
between word-vector maps and classification possibility.
%This in itself is a step forward; ultimately the current QNLP methods avoid 
%robustness and scalability, which is the primary focus of using quantum state representations at all. 
So far we have been able to achieve 100\% accuracy on the Lambeq text classification dataset,
and 62\% average accuracy on a more realistic dataset of 50 movie reviews.

It is clear that a quantum encoded 
feature map can classify text sets in simulation, but so far for small and carefully targeted samples.
The most obvious next steps include running these workflows on QPU resources: these experiments are underway
(and are expected to be included when this paper is presented).

The focus on binary classification is another simplification that has made this work more tractable
initially but is clearly a limiting restriction. Standard methods for extending 
SVMs to $m$ classes include building $m$ one-vs-rest classifiers, or other collections of 
pairwise classifiers that can be combined to make $m$-way decisions. This may be a natural next step
for quantum classification work. A more imaginative conjecture would be that the geometry of
particular quantum feature spaces lends itself to new opportunities for separating the space into $m$
topological components, an avenue we have yet to pursue.

As stated, the text preprocessing was so far simple, and typically carried out during 
the initial word vector encoding step. The impact of these choices on results has so far not been 
investigated. A more ambitious proposal would be to find ways to perform more of the distributional
vector encoding on quantum computers, as begun by \cite{lorenz2021qnlp,widdows2022near}.

There are many interesting avenues to pursue that can push towards quicker 
improvement of QNLP and quantum machine learning methods. This work addresses primarily space-efficiency, and aims to incentivize work 
regarding  text encoding schemes, robust state preparation methods and algorithms for working with text data. Moving forward, 
investigations are in progress to determine patterns between feature map construction and its effect on text data mapping, with error 
analysis both in and out of simulation. Further work to develop a complete view of the trade-off between space, simplicity, and processing 
time is necessary to make an objective opinion on the potential of classification in QNLP.

%%%%
\bibliography{bib/ionq}
\bibliographystyle{IEEEtran}

\end{document}